\documentclass{egpubl}

\ConferenceSubmission   
\ifpdf \usepackage[pdftex]{graphicx} \pdfcompresslevel=9
\else \usepackage[dvips]{graphicx} \fi

\PrintedOrElectronic

\usepackage{t1enc,dfadobe}

\usepackage{egweblnk}
\usepackage{cite}

\title[Aquanims area-preserving animated transitions]%
      {Aquanims: Area-Preserving Animated Transitions in Statistical Data Graphics based on a Hydraulic Metaphor}

\author{Michael Aupetit \thanks{ email: maupetit@hbku.edu.qa}\\ %
        \scriptsize Qatar Computing Research Institute, HBKU, Doha, Qatar}

\begin{document}


\maketitle

\begin{abstract}
  We propose "\textit{aquanims}" as new design metaphors for animated transitions that preserve displayed areas during the transformation. Animated transitions are used to facilitate understanding of graphical transformations between different visualizations. Area is a key information to preserve during filtering or ordering transitions of area-based charts like bar charts, histograms, tree maps or mosaic plots. As liquids are incompressible fluids, we use a hydraulic metaphor to convey the sense of area preservation during animated transitions: in \textit{aquanims}, graphical objects can change shape, position, color and even connectedness but not displayed area, as for a liquid contained in a transparent vessel or transferred between such vessels communicating through hidden pipes. We present various \textit{aquanims} for product plots like bar charts and histograms to accomodate changes in data, in ordering of bars or in number of bins, and to provide animated tips. We also consider confusion matrices visualized as fluctuation diagrams and mosaic plots, and show how \textit{aquanims} can be used to ease the understanding of different classification errors of real data. \\

\begin{classification} 
\CCScat{Computer Graphics}{I.3.3}{Picture/Image Generation}{Line and curve generation}
\end{classification}

\end{abstract}

\section{Introduction}

Visualization can be used to discover new information or to communicate findings. Discovery often requires changing points of view, filtering data, modifying the models, selecting different metaphors. Communicating requires explaining the process that transformed the data into insights. 
In both cases, transformations occur between the visualizations. Animated transitions are key to keep the user understanding the displayed information during these phases, linking seemingly different representations. 

A recent survey \cite{ChevalierRPCH16} showed that animated transitions can help keep the information in context, be used as a teaching aid, improve user experience, support the visual discourse or enhance data encoding for discovery. However animations are not always guaranteed to improve the visualization performances \cite{HeerR07} \cite{ChevalierDF14}. For instance too many objects moving can be disturbing and prevent tracking the objects of interest, or occlusions may hide some objects during the transition. 

In statistical graphics, data exploration and understanding are supported by specific visual metaphors like bar charts, scatterplots, heatmaps or node-link diagrams. We focus on area-based charts which use the size channels to encode the primary data. In these charts, the area of a graphical element is proportional to the underlying count, proportion or probability.
Many of them have been encompassed in the product plot framework \cite{WickhamH11} related to statistics. Preserving area during the transition is key to comply with the congruence principle \cite{TverskyMB02}, i.e.to maintain graphics validity during the transition, and semantic-syntactic mapping using similar metaphors for similar transitions. Minimizing occlusion and maximizing predictability are two other important laws of animations \cite{HeerR07}.

However, animated transitions between different area charts distort these areas: complete or partial occlusion modify the perceived area; very standard linear interpolation between rectangles with different aspect ratio change their area; changing the number of bins of an histogram while maintaining their total area through a continuous transition is not trivial too.

In this work, we propose a physical metaphor to guide the design of area-preserving animated transitions. It is based on hydraulics. First, because liquids are incompressible fluids that are likely to convey the sense of volume preservation projected as areas on our retina. Second because we are very used to manipulate liquids in our everyday life: pouring milk in a glass, filling a sink, emptying a plastic bottle, transferring water between buckets or in a water tank, drinking a soda with a straw... not to mention experimental physics at school or in professional environments. We call \textit{\textit{aquanims}} the animated transitions based on the proposed hydraulic metaphor.   

In section 2, we review the related work. In section 3 we present the animation principles based on the hydraulic metaphor and how they are related to the guiding principles of good animations. In section 3, we present the hydraulic metaphor and the \textit{aquanims} fundamental principles. In section 4 we describe the building-blocks to be used in section 5  to create \textit{aquanims} for various area-based charts. In section 6, we use it to display different facets of a confusion matrix to analyze the results of a classification problem. Before we discuss in section 7 and  conclude in section 8.

\section{Related work}

The reference work of \cite{HeerR07} investigates the effectiveness of animated transitions between statistical data graphics and connects the guiding principles of good visualizations to the ones for good animations proposed by Tversky \cite{TverskyMB02}. In particular, transitions should avoid occlusion; maintain graphical encoding at each and every interpolation state; use same semantic-syntactic mapping across the different visualizations; make change only of the necessary syntactic features (expressiveness) to make the animation best support (efficiency) the communication of the semantic changes; use staging when transition is complex, and make it short. 
These authors also propose a typology of animations, in this work we explore all the types except the \textit{view} and \textit{substrate} transformation ones.
Wickham and Hofmann \cite{WickhamH11} proposed the \textit{product plot} framework which encompasses many area-based charts. We use this framework to design the animated transition between a fluctuation plot and a mosaic plots used to visualize confusion matrices. 

Using metaphors is common in visualization \cite{ZiemkiewiczK08} to influence the representation of information in the mind, like Furna's generalized Fisheye views for instance \cite{furnas86}. \textit{Visual Sedimentation} is a design metaphor inspired by the physical process of sedimentation. It is especially suited to represent data stream. New data enter the visualization as bubbles attracted toward buckets, and progressively accumulate and create strata in form of color-coded area charts which summarize past data and progressively  compress the information stream. We are not aware of a metaphor based on liquids or hydraulics for information visualization.

Animated transitions have been used to explain and support discovery with statistical graphics. For instance ScatterDice \cite{ElmqvistDF08} uses a pseudo rotation of the axes to navigate a scatterplot matrix. Pairing a complex graphics to a simpler one with animated transitions has been proposed in \cite{RuchikachornM15} to help learn how to use complex graphics. In LineUp \cite{GratzlLGPS13}, multiple rankings represented as bar graphs can be compared and reordered. Animation is used for the reordering part but when bar charts are stacked together by the user, the transition is not animated. A vertical fish-eye lens with animated transition is also proposed to enlarge bars otherwise too thin to be visible when there are too many of them.


\section{\textit{aquanims}}

\subsection{A hydraulic metaphor}

Hydraulics is the branch of science and technology concerned with conveying liquids through pipes and channels, and to use it as a source of mechanical force or control. 
Typical hydraulic technology connects cylinders, tanks and pumps with pipes. As liquids are nearly incompressible, they keep their volume constant in the system, and they can transmit forces between distant containers connected by non-elastic pipes. Liquids have a mass so are subject to gravity forces, leading to equal surface level equilibrium in possibly distant but communicating vessels, as well as in natural artesian wells.
Now we detail the analogy between graphics and hydraulics and derive the fundamental  principles of \textit{aquanims}.

\subsection{Fundamental principles of \textit{aquanims}}

As tanks and vessels in hydraulic systems, area charts have containers like bars or tiles of which the area is measured. We will focus on bars and tiles of rectangular shape with axis-parallel edges in the sequel, although the analogy is not limited to these shapes and orientations and could be used for pie charts for instance. Bars and tiles are usually the graphical objects whose area must be preserved as they encode the primary data. 

However in our physical analogy we consider that the equivalent to a bar or a tile is actually the liquid which takes the form of its graphical container. By so doing we transfer the semantic of the area of the graphical container to the one of the graphical fluid it contains. So we can change the shape of a set of containers, move the liquid from one container to another, or even let it be shared between several containers without changing its total volume. 

Thus we derive the very first fundamental principle underlying \textit{aquanims}: 

\textbf{First principle: the liquid encodes the data.} Data are mapped to the fluid  content rather than to its container. 

The second fundamental principle is obvious from the physical analogy: 

\textbf{Second principle: the liquid volume is constant}, so is its area in the graphical representation. Pipes and containers can change the liquid's shape but cannot change its total area. 

As a \textbf{main design guiding principle:} \textit{aquanims} should be designed so as to  emphasize the hydraulic metaphor evoking liquid-coded data (First principle), in order to increase the perception of area invariance (Second principle).

As a consequence, because the container is not encoding data directly, the liquid itself must bear some identification mark. We use the color channel for this sake (texture could be used also) to fill the rectangles or color its edges.

Now we focus on the \textit{aquanims} building-blocks that simulate hydraulics.

\section{\textit{aquanims} building-block} 

\begin{figure*}
  \centering
  \mbox{} \hfill
  \hfill
  \includegraphics[width=\linewidth]{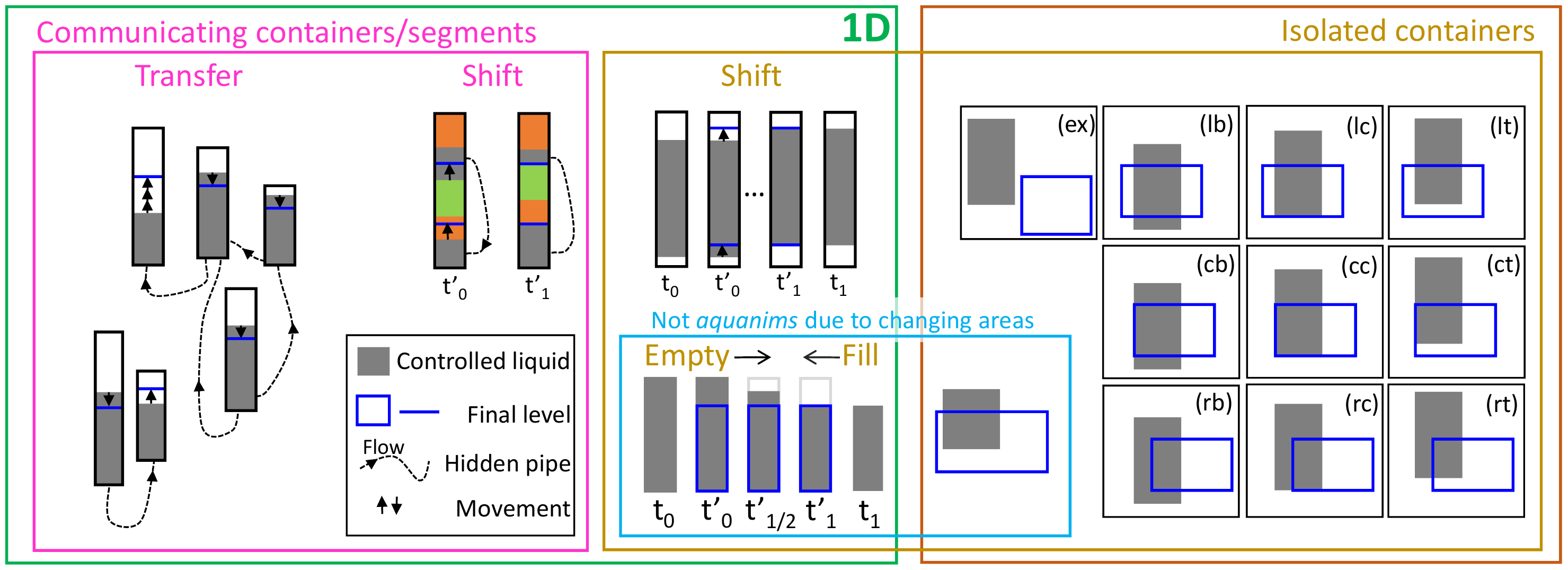}
  \hfill \mbox{}
  \caption{\label{fig:buildingblocks}%
           Building-blocks of \textit{aquanims}. Filled gray rectangles show the initial state, and blue empty rectangle the final state of the animations. Both rectangles have the same area except empty/fill animations in the blue frame depicting non \textit{aquanims} per se but useful blocks to build the communicating containers and segments (purple frame) aquanims. All transitions can be reversed. Liquid can move along the container axis (1D green frame) or in both dimensions (2D brown frame). Isolated containers (dark yellow frame) have no hidden pipes. The liquid is either shifted along the container axis (1D) or reshape (2D) by reshaping the container itself as described in more detailed in figure \ref{fig:reshapedetails} and \ref{fig:reshapecases}. }
\end{figure*}

We illustrate the different building-blocks that support the design of \textit{aquanims} in the figure \ref{fig:buildingblocks}.

We distinguish between changing the position (translation) or size of a container in the underlying space, or the view (pan, zoom) and scale (rescaling) of the underlying space itself, and changing the container shape (resize, reshape). The latter involved \textit{aquanims} while the former do not. To comply with animation principles of congruence and consistent semantic-syntactic mappings, we distinguish graphically pan, zoom and rescaling which involve animation of the axes and background and associated changing displayed size and position of all the containers and liquids they contain, while translation, resize or reshape occur with no background changes. \textit{aquanims} modify size, shape and fill using additional graphical elements evoking the hydraulic metaphor and area-preserving property.

\begin{figure*}[htb]
  \centering
  \includegraphics[width=0.8\linewidth]{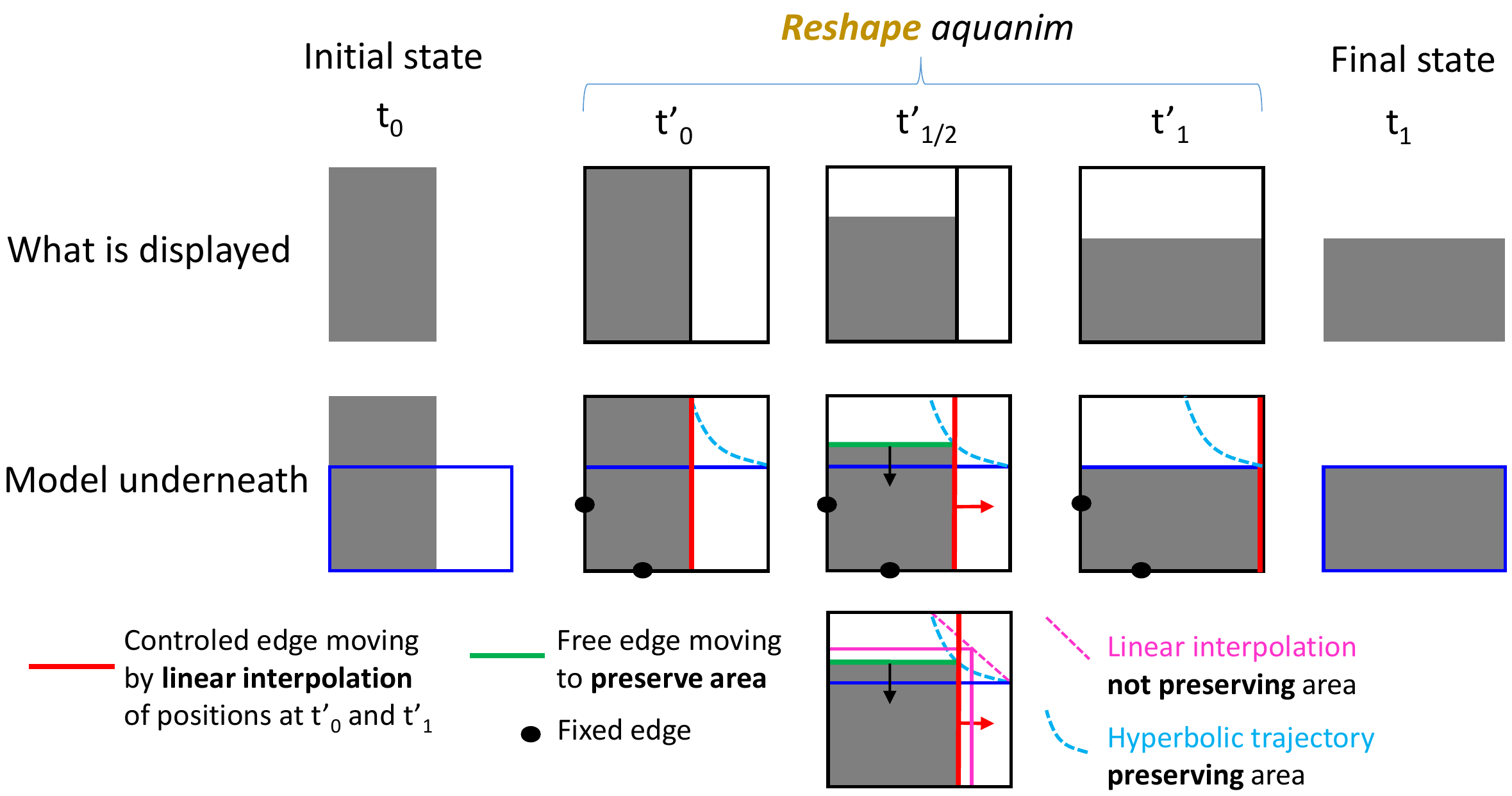}
  \caption{\label{fig:reshapedetails}%
Design details of Reshape \textit{aquanims} following a hydraulic cylinder metaphor with cylinder in black, piston in red and free liquid surface in green. Notice how the free surface is bounded by the piston and the piston bounded by the cylinder (see also figure \ref{fig:reshapecases}).
Top row shows the aquanim as displayed. A first step add the bounding box of both initial and final state rectangles and the side which is controled (middle vertical line, red color in the second row) is interpolated linearly between initial ($t'_0$) and final states ($t'_1$). Then this elements are removed to render the final state.  The underneath model (second row) show that here two sides of the inital rectangle are fixed (black dots), one is directly controlled (red line) and one is moved (green line) under the preserving-area constraint. The corners of the reshaped rectangle follow hyperbolic trajectories (light blue dashed lines). The bottom row shows that linear interpolation (purple box and dashed line) between initial and final positions of the vertices does not preserve area. }
\end{figure*}

\begin{figure}[htb]
  \centering
  \includegraphics[width=\linewidth]{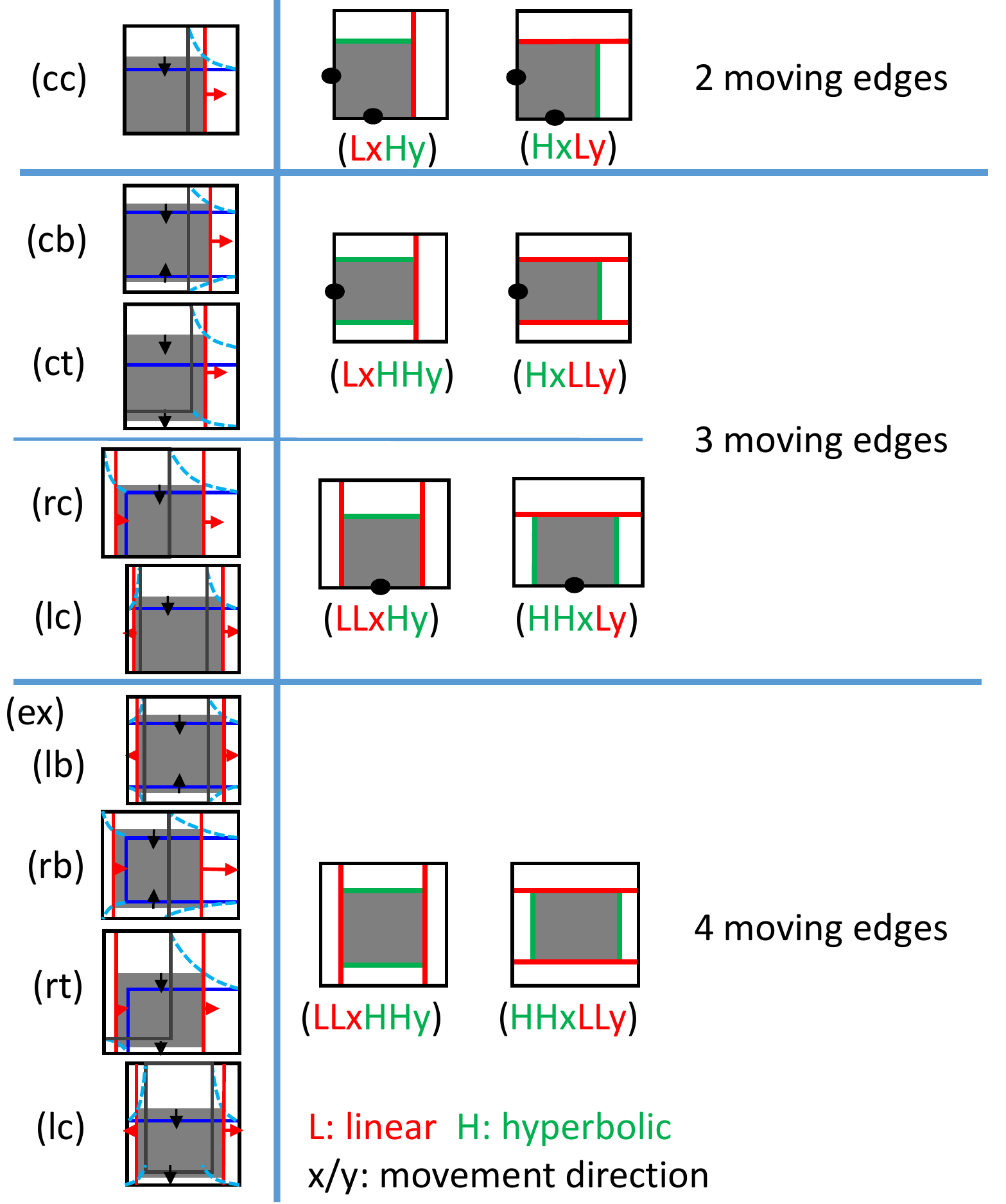}
  \caption{\label{fig:reshapecases}%
Typology of reshape \textit{aquanims}. 
The legend and left codes correspond to the ones given in figure \ref{fig:buildingblocks}. The left column shows the detailed intermediary state interpolated between initial and final states for each of these cases except for (ex) which can be similar to any of the ones of the bottom row except the dark gray and blue boxes do not cross. The middle column shows a schematic representation of the detailed cases showing which edges are linearly interpolated (red) or follow a hyperbolic trajectory (green) to preserve area. The two sub-columns show alternative designs where the piston moves either along x (left) or y (right) axes. L and H letters code for the linear and hyperbolic interpolations respectively, used along the axes. Double LL and HH mean two edges moving in that way. The right column indicates the number of edges moving. We conjecture that the best aquanims in terms of perception of area-preservation are the ones for which a minimum number of edges are moving (first row) or at most a single edge is moving with a hyperbolic interpolation (cases HxLLy and LLxHy preferred in case of 3 moving edges)}
\end{figure}

\subsection{Isolated container: empty/fill}

A container can be \textbf{filled or emptied}. We represent the container by an empty rectangle and the liquid by a partial filling of it as a smaller rectangle having 3 sides shared with the container. The animated transition uses a \textit{slow-in/slow-out} speed  based on a cubic basis function: $u(t)=3t^2-2t^3$ for a regular sequence of values $t\in[0,1]$.

In order to comply with the predictibility principle of animated transitions, for \textit{emptying} animations, the final level of the liquid in the container is indicated by a \textit{target} line segment throughout the transition. The initial level being explicitly given by the container edge. For \textit{filling} animation, the edges of the final container is displayed and progressively filled by the liquid, while the initial level is displayed with a line segment as a reminder. Notice that in the figure \ref{fig:buildingblocks} we use bright colors to emphasize edges for explanations but when applying aquanims, colors shall be adapted to the palette used.   

This  \textit{filling/emptying} animation is identical to the one of standard progression bars, or volume bars in digital equalizers. 
The initial and final line segments help the user perceive the transition process, remember the initial state and predict the final one. 

However this animation alone violates the second principle of \textit{aquanims} that the liquid area must be preserved, so this building-block is not an \textit{aquanim} \textit{per se} but it enables the construction of aquanims based on communicating containers or segments (see sections \ref{sec:communicatingcontainerstransfer} and \ref{sec:communicatingsegmentsshift}.

\subsection{Isolated container: shift} 

The liquid in a container can be \textbf{shifted}, \textit{i.e.} translated along the container axis, so the rectangle representing the liquid is translated while the container remains fixed in the underlying space (Figure \ref{fig:buildingblocks} in intersection of yellow and green frames). This animation is area-preserving so is an \textit{aquanim}.

\subsection{Isolated container: reshape} 
\label{sec:isolatedcontainerreshape}
A container can be \textbf{reshaped} so the position of more than one edge is changed (2D brown frame in the figure \ref{fig:buildingblocks}.
Compared to the filled/emptied animation, at least 2 edges oriented in orthogonal directions are changing position. As a consequence, the increase in one dimension must be compensated for by a decrease in the other to preserve area (Figure \ref{fig:reshapedetails} top).

We use a hydraulic cylinder metaphor for these \textit{aquanims} (Figure \ref{fig:reshapedetails} center): a bounding box encompassing both the initial and final rectangles is displayed to represent the cylinder; The moving edges under control are prolonged toward the edges of the cylinder, to represent pistons; And remaining edges are either fixed in contact to the cylinder, or free edges joining two pistons or a piston and the cylinder.

It must be noticed that a linear interpolation between initial and final positions of the rectangle's vertices does not preserve area (Figure \ref{fig:reshapedetails} bottom).
We linearly interpolate (L code) one of the dimensions while we compute the remaining one using a hyperbolic interpolation (H code) which maintains the area of the rectangle invariant during the transition. 

The figure \ref{fig:reshapecases} show that the $10$ cases described in the figure \ref{fig:buildingblocks} right, reduce to 4 cases ignoring symmetry between x and y rectangle axes: $L.H.$, $L.HH.$, $LL.H.$ and $LL.HH.$ where $L$ and $H$ denote linear and hyperbolic interpolations respectively, and are duplicated when two edges are animated this way.
We assume that linear interpolation is more predictible than hyperbolic one, and so as an efficient design guiding principle, the lower the number of moving edges ($L.H.$ preferred), and the lower the number of $H$-type ones among them ($LL.H.$ preferred to $L.HH.$), the better the perception of the hydraulic metaphor and so of the area-preserving property. In cases where more than 3 moving edges are required (bottom rows), we might advise to use a staged animation adding a translation of the rectangle before or after one of the cases (.c) or (c.).

The general formula for these area-preserving animations are:

$l_i(t)=(1-u(t)) l_i(0) + u(t) l_i(1)$ and $ H(t)=\mathcal{A}/L(t)$

with $t\in[0,1]$. $l_i$ is the position of the piston edge $i$ along the axis orthogonal to it and controlled directly by the slow-in/slow-out speed transition. $L(t)=|l_{i_1}(t)-l_{i_2}(t)|$ and $H(t)=|h_{j_1}(t)-h_{j_2}(t)|$ are the dimensions of the rectangle at step $t$. $h_j$ is the position of the free edge $j$ along its orthogonal axis and controlled by the area-preserving condition regarding the area $\mathcal{A}=L(0)H(0)=L(1)H(1)$ of both initial and final rectangles. When two edges follow a hyperbolic interpolation, we can find the $h_{j_!}(t)$ and $h_{j_2}(t)$ values by adding a constraint to center the rectangle along the $H$-animated dimension:
$c(t)=(1-u(t)) c(0) + u(t) c(1)$

where $c(t)=(h_{j_1}(t)+h_{j_2}(t))/2$.

\subsection{Communicating containers: transfer} \label{sec:communicatingcontainerstransfer}

Two or more containers can be connected by invisible pipes allowing continuous  liquid transfer between them during the animation (see Figure \ref{fig:buildingblocks} purple frame). Any amount of liquid which disappears from one container must be distributed among the other ones, so the total sum of variation of the connected containers is zero, maintaining the total amount of liquid at a constant value.

We have this formula between variation $\delta_k$ of levels within connected containers of respective width $w_k$:

$$\sum_k w_k \delta_k=0$$

Assuming fixed-width containers like in bar charts and histograms, if the initial $\lambda_k(0)$ and target $\lambda_k(1)$ levels in each containers are such that total area $\mathcal{A}=\mathcal{A}(0)=\sum_k w_k \lambda_k(0)=\sum_k w_k \lambda_k(1)=\mathcal{A}(1)$ is constant for these two states, then for any linearly interpolated state $\lambda_k(t)=(1-u(t))\lambda_k(0)+u(t)\lambda_k(1)$, the area is preserved $\mathcal{A}(t)=\sum_k w_k \lambda_k(t)=\mathcal{A}$.

So linear interpolation with possible slow-in/slow-out speed or any other speed scheme ensures the second principle of \textit{aquanims}.

\subsection{Communicating segments: shift} 
\label{sec:communicatingsegmentsshift}
In a single container, two or more segments of the controlled liquid separated by segments of other liquids can be connected by  hidden pipes following the same rules as communicating containers (see Figure \ref{fig:buildingblocks} purple frame). The liquid transferred in the pipe changes the local segment area but preserves global area. The other liquids are shifted along the container. 

In the next section we will present different \textit{aquanims} we designed based on these building-blocks to animate histograms, and bar graphs.

\section{\textit{aquanims} design case study}

Product plots \cite{WickhamH11} is a framework which encompasses many area-based charts.
We picked some of the most well known ones to illustrate the design of \textit{aquanims}. We present the case of vertical bars but it is trivial to adapt the animation for horizontal ones.
Letters refer to the figure mentioned in each section. We developed the animations using R programming language and Shiny framework for web interfaces, with the \textit{ggplot2} and \textit{animation} R packages. 

\subsection{Histograms}

A histogram is a type of bar chart with continuous x and y scales, where bars are adjacent to evoke the continuous nature of the x-axis (if vertical bars). We consider the case of equal width bars, which define contiguous intervals on the x-axis. The area of one bar accounts for the proportion of data falling in that interval or bin, the total area of the bars being 1.
All our animations are staged, starting and ending with a pan and zoom phase, at first to adapt the size of the plot so that the subsequent \textit{aquanim} fits entirely within the graphic area, then at the end to maximize the size of the plot in this area.

\subsubsection*{Adapt to changing data}
\label{sec:histoadaptchangingdata}

\begin{figure}[htb]
\centering
\begin{tabular}{ccc}
\includegraphics[width=0.3\linewidth]{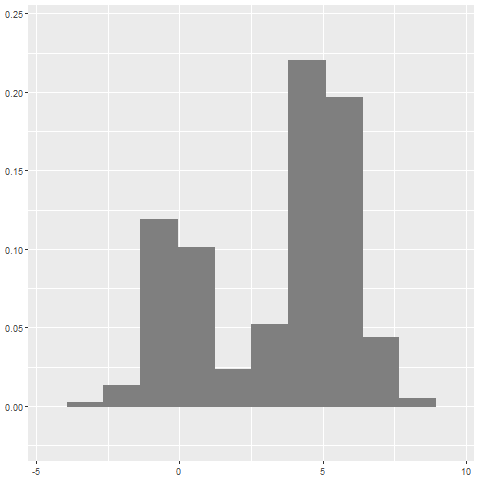}
&\includegraphics[width=0.3\linewidth]{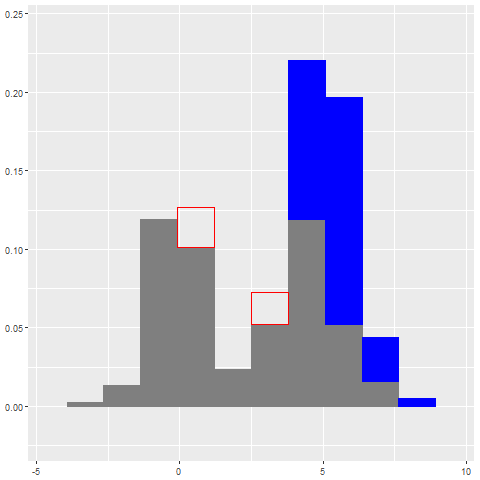}
&\includegraphics[width=0.3\linewidth]{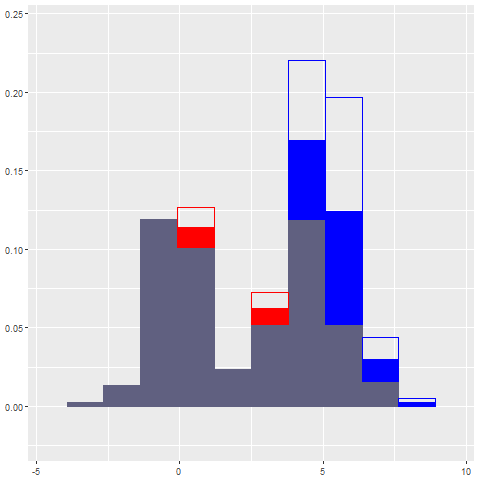}\\
(a)&(b)&(c)\\
\includegraphics[width=0.3\linewidth]{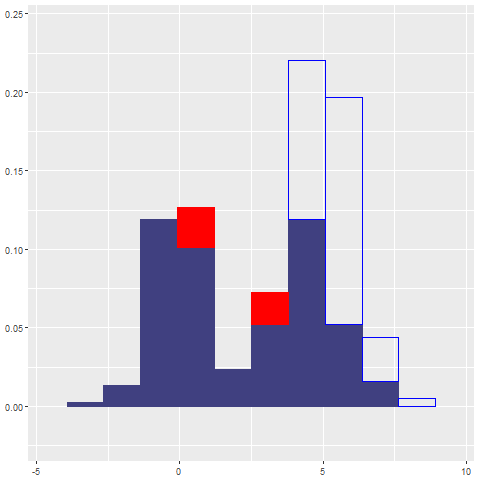}
&\includegraphics[width=0.3\linewidth]{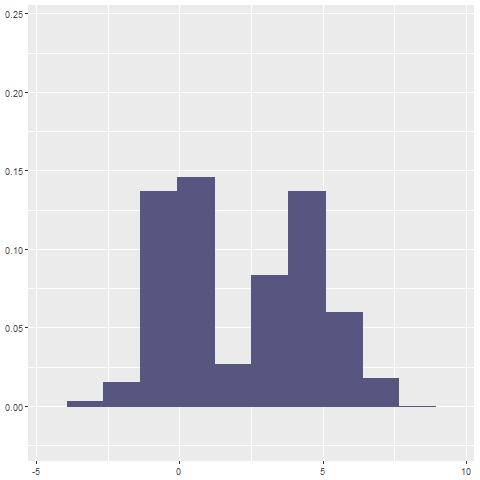}
&\includegraphics[width=0.3\linewidth]{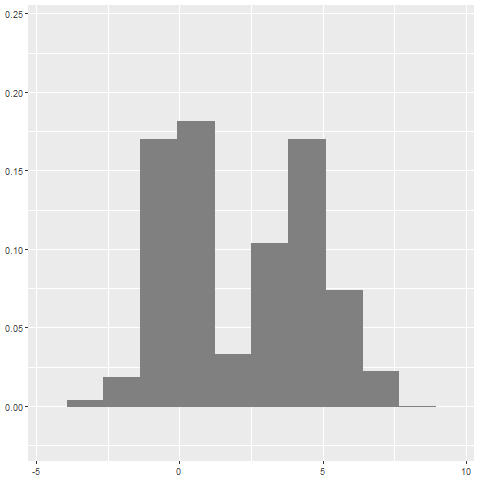}\\
(d)&(e)&(f)\\
\end{tabular}
\caption{\label{fig:adapchangingdata}%
\textit{Aquanim} of a histogram for adding (red) and deleting (blue) data values (b). In (c),(d) and (e) the bluish color of the bars indicates that the total density is lower than 1 (summing  up added and removed data), the aquanim resizes the bars  (e) to finally compensate for that lack of data  (f).}
\vspace{-.5cm}
\end{figure}

We consider the case of adapting an histogram to data filtering where some data can appear and some other disappear (Figure \ref{fig:adapchangingdata}). The area of the histogram must remain constant.

We use a plain red hue for more data in a bar, and plain blue for less data in a bar (b,c,d). The red rectangles are to be filled by red color during the \textit{aquanim} using the single container fill-block animation and  the blue rectangles have to be emptied using the empty-block animation. During the filling, we make the gray color of all the bars reddish or bluish (c,d)  to evoke the over or under pressure in terms of total area which transiently becomes higher or lower than 1 depending on the overall surplus or lack of data. Then we rescale  all the bars (d,e,f) so the area comes back to unity and the reddish or bluish color disappears (f).

\subsubsection*{Adapt to changing number of bars}

\begin{figure}[htb]
\centering
\begin{tabular}{cc}
\includegraphics[width=0.4\linewidth]{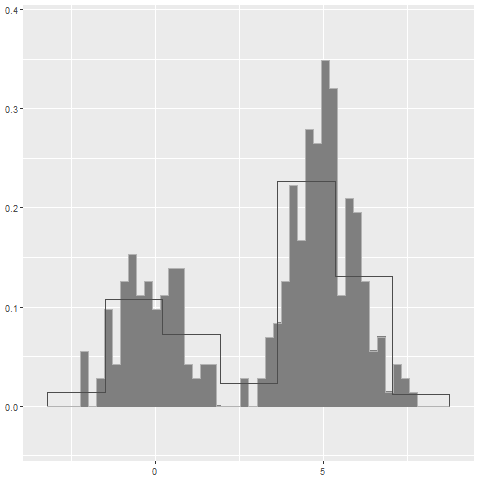}
&\includegraphics[width=0.4\linewidth]{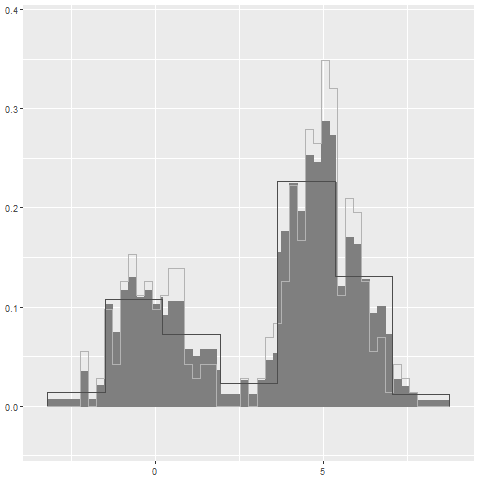}\\
(a) &(b)\\
\includegraphics[width=0.4\linewidth]{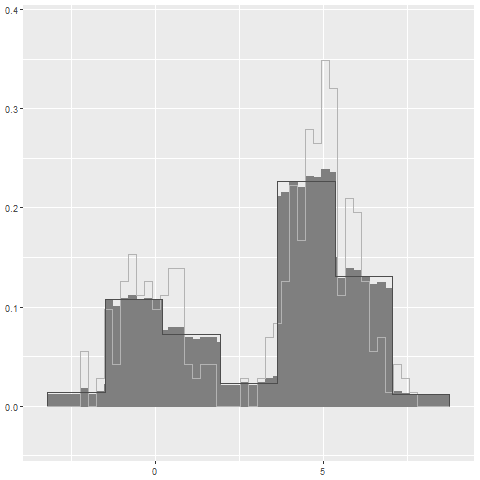}
&\includegraphics[width=0.4\linewidth]{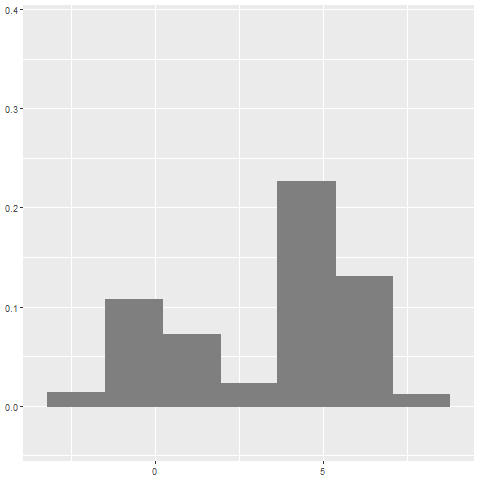}\\
(c)&(d)\\
\end{tabular}
\caption{\label{fig:adapchangingnumbins}%
\textit{Aquanim} for a decreasing number of histogram bins. From (a) to (d), levels of intersecting bins are linearly interpolated. Reading from (d) to (a) shows the \textit{aquanim} for increasing number of bins.}
\vspace{-.5cm}
\end{figure}

Histograms are statistical models which approximate the underlying data density. There is no straightforward way to select the best number of bars: too many it becomes spiky, and to few all the details are smoothed, but the user might want to test different numbers so we propose an \textit{aquanim} to smoothly transition between the results.

Adding bars could be done one by one, but would require to decide arbitrarily where to add the new bar (left, right, middle?). It is also not consistent with the semantic-syntactic mapping:  bars are discretizations of a continuous axis, so the incremental addition of whole bars hides the continuous nature of the underlying data. Adding the bars from one side incrementally also generates a sliding effect, giving the erroneous impression of a zoom out or rescaling of the x-axis. The same issues occur when removing bars one by one.

We use the communicating containers transfer block (Figure \ref{fig:adapchangingdata}). We superimpose the contours of the original (light gray) and target (dark gray) histograms. Thin bars can be merged into large bars or a large bar be split into the thin bars it overlaps. A linear interpolation (a,b,c) between initial and final levels within each bar of the intersection is used for the transition. We also tried a more physically realistic approach, where we made adjacent bars as communicating vessels, and set up an adaptive formula that makes each level converging toward the average level of its two neighbors, normalizing the area at each step. The result can be seen in the supplemental material. How realistic the liquid dynamic should be to optimize the area-preserving perception has not been investigated. 
 
\subsubsection*{Show animated tip}

\begin{figure}[htb]
\centering
\begin{tabular}{cc}
\includegraphics[width=0.4\linewidth]{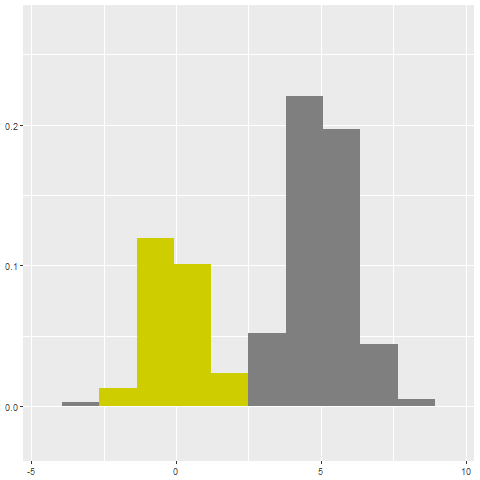}
&\includegraphics[width=0.4\linewidth]{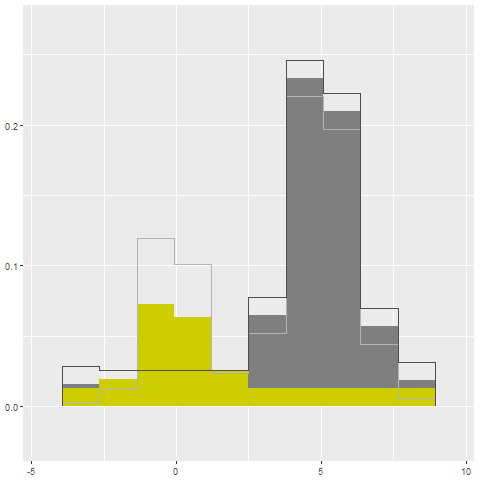}\\
(a) &(b)\\
\includegraphics[width=0.4\linewidth]{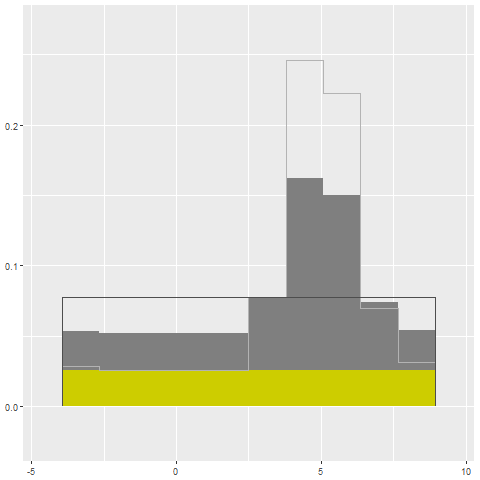}
&\includegraphics[width=0.4\linewidth]{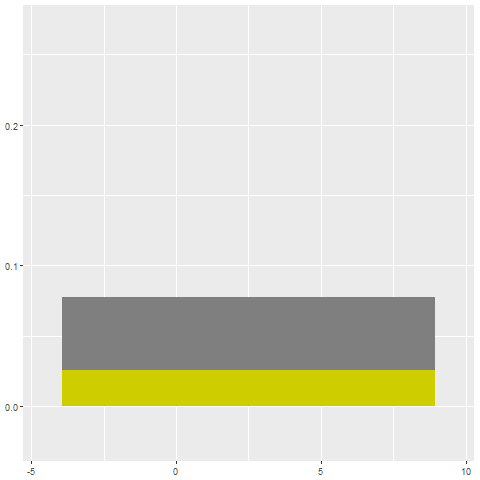}\\
(c)&(d)\\
\end{tabular}
\caption{\label{fig:animatedtip}%
\textit{Aquanim} tip to show the proportion of the total represented  by the selected bars (yellow). The animation starts in (a) up to (d) then after a pause or on user action, come back from (d) to (a).}
\vspace{-.5cm}
\end{figure}

The idea here is to show the user what proportion of the total data a set of bars represents  (Figure \ref{fig:animatedtip}). As all the bars may have different heights, this is not easy to evaluate visually from the original histogram. 

We color the selected bars in yellow to distinguish them from the rest (a). We empty the yellow liquid into the bottom of all the bars, raising all them up to preserve the total area (b). We then release the gray liquid to equalize the level of the grey bars (c), so the final graphic is a vertical stacked bar with a yellow segment at the bottom and a gray one at the top (d), where the sought proportion of selected data is directly visible. This \textit{aquanim} tip can be designed with several colors to several sets of bars selected by some clustering algorithm for instance.

\subsection{Bar charts}

Bar charts are bars that stand on a categorical or discrete ordinal axis, and whose height maps a continuous variable. Bars can be segmented to distinguish levels of some independent category. These segments can be stacked on top of each others or grouped side by side. Heer and Robertson studied different transitions between stacked and grouped bars in \cite{HeerR07}.

\subsubsection*{Stacked bars vertical reordering}

\begin{figure}[htb]
\centering
\begin{tabular}{cc}
\includegraphics[width=0.4\linewidth]{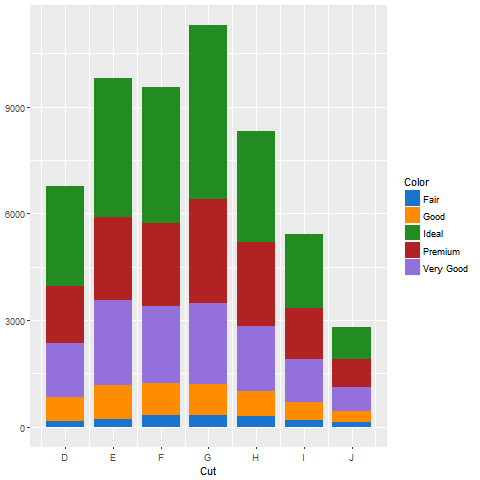}
&\includegraphics[width=0.4\linewidth]{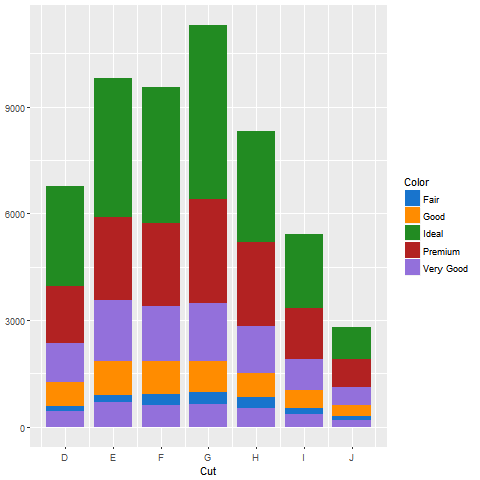}\\
(a) &(b)\\
\includegraphics[width=0.4\linewidth]{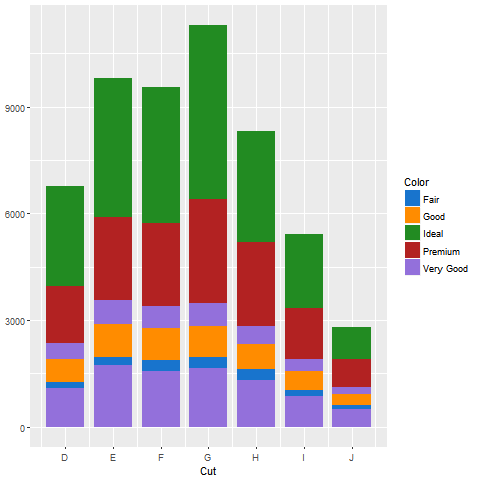}
&\includegraphics[width=0.4\linewidth]{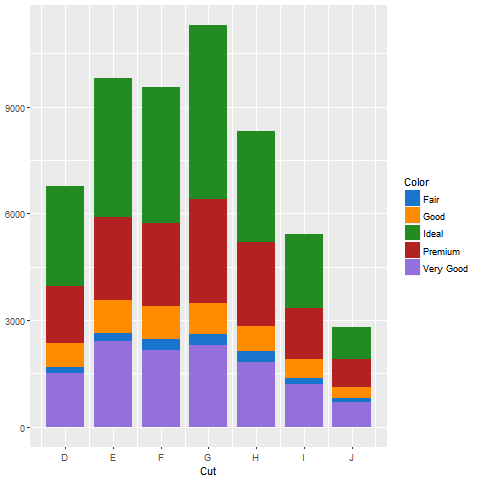}\\
(c)&(d)\\
\end{tabular}
\caption{\label{fig:aquanimstackbarvertreorder}%
\textit{Aquanim} of segments in vertical bars to move the selected ones (magenta) at the bottom.}
\vspace{-.5cm}
\end{figure}

We consider the typical task of reordering the segments of stacked bars, as only the bottom ones along the x-axis benefit from this common straight line to ease the comparison of their amount across the levels of the categorical variable mapped to this axis (Figure \ref{fig:aquanimstackbarvertreorder} (a,d)).

We use the communicating-segments-shift block directly, and apply it to all the bars at once. The liquid in the segments of the selected color-coded level progressively fills the bottom part of their bar (b,c), shifting up the intermediate segments until the original ones vanish from their initial position (d). The selected segments are split, preserving their total area in each bar while avoiding occlusion that standard animated translations would generate.  

\subsubsection*{Stacked bars horizontal reordering}

\begin{figure}[htb]
\centering
\begin{tabular}{ccc}
\includegraphics[width=0.3\linewidth]{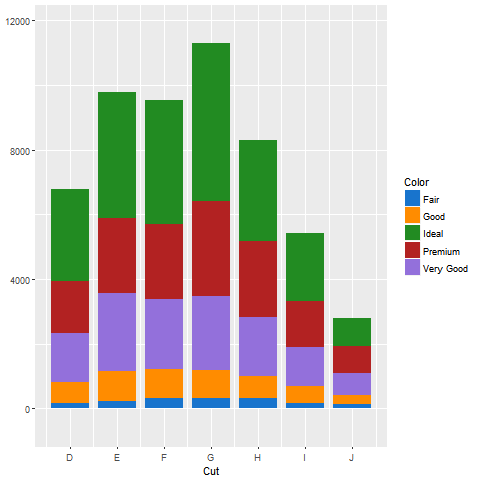}
&\includegraphics[width=0.3\linewidth]{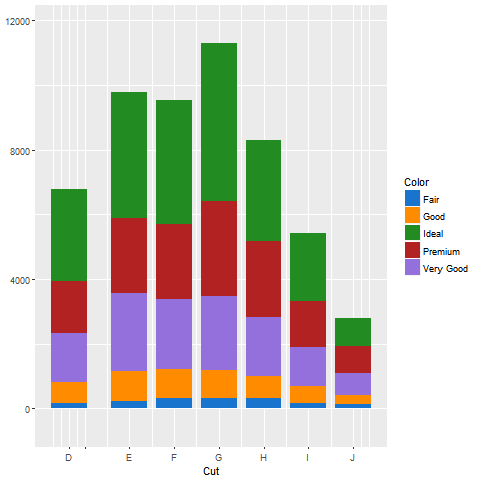}
&\includegraphics[width=0.3\linewidth]{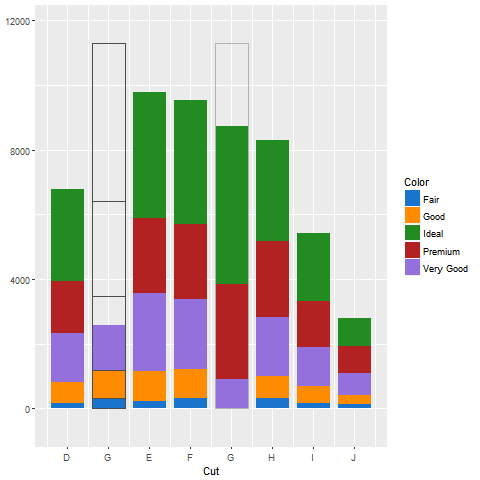}\\
(a) &(b)&(c)\\
\includegraphics[width=0.3\linewidth]{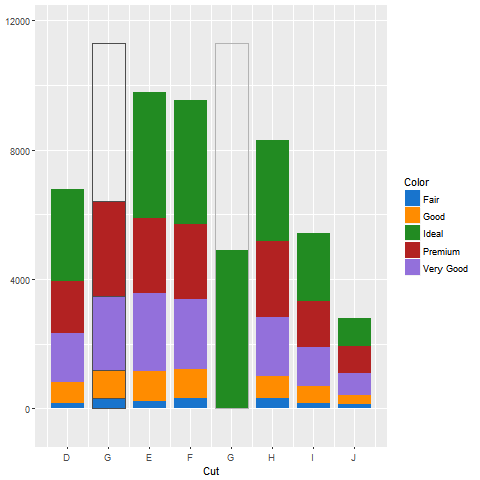}
&\includegraphics[width=0.3\linewidth]{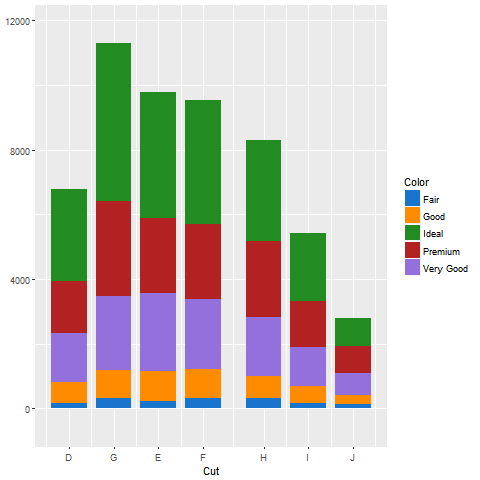}
&\includegraphics[width=0.3\linewidth]{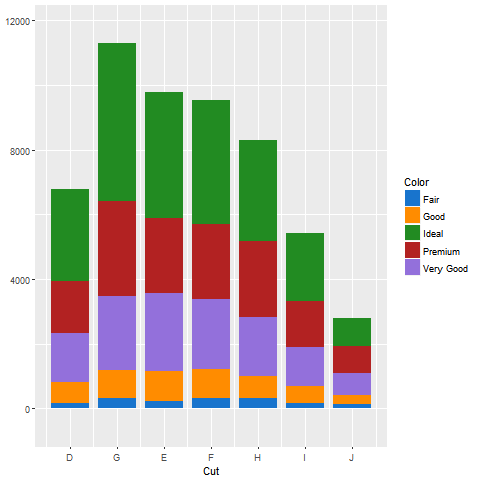}\\
(d)&(e) &(f)\\
\end{tabular}
\caption{\label{fig:aquanimstackbarhorizreorder}%
\textit{Aquanim} to reorder vertical bars. The bar G is moved between D and E using the communicating- containers-transfer block.}
\end{figure}

We propose an \textit{aquanim} to translate a bar  from an initial to a final position without occlusion (Figure \ref{fig:aquanimstackbarhorizreorder}). 

We first create an empty space at the final position between levels D and E (a,b), with the same label G as the initial. Then we use the communicating-containers-transfer block, progressively emptying the original bar segment by segment starting from the bottom, and filling the destination bar at the same time by the liquid coming from the initial one (c,d). Then we remove the space let by the empty bar at the initial position (e,f).

Here again the split of the transferred segments between two bars avoid  occlusion and let invariant the visible area of each segment during the transition. The same animation could be used to transfer bars in a grouped bar chart. Transferring multiple bars at the same time might be confusing, a staggered animated transition could be used in that case \cite{HeerR07}.

\section{Case study: animation of a confusion matrix}

We consider classification results obtained from an automatic classifier.
Access to situation-sensitive data through social media networks provided images of damages after natural disasters. Rapid damage assessment is important for many humanitarian organizations. A Convolutional Neural Network has been used to classify these images into 3 categories: \textit{None}, \textit{Mild} and \textit{Severe}. CITE PAPER OF DAT

\subsection{Confusion matrices in classification}

\begin{table}
\begin{center}
\begin{tabular}{|l|c|c|c|}

\hline
& None & Mild & Severe\\
\hline
None & 1458  & 48 &  78\\
\hline
Mild & 205 & 102 & 144\\
\hline
Severe &  85 &  34& 1666\\
\hline
\end{tabular}
\end{center}
\caption{ \label{tab:confmat} \textbf{Confusion matrix} The data table representation of a confusion matrix gives all classification results details but fails to give an overview and focus rapidly on the anomalies. The rows are predicted classes, and the columns are observed ground truth classes. For instance there are $85$ images with no damage but predicted showing severe ones.}
\end{table}

\begin{figure}[htb]
  \centering
  \includegraphics[width=\linewidth]{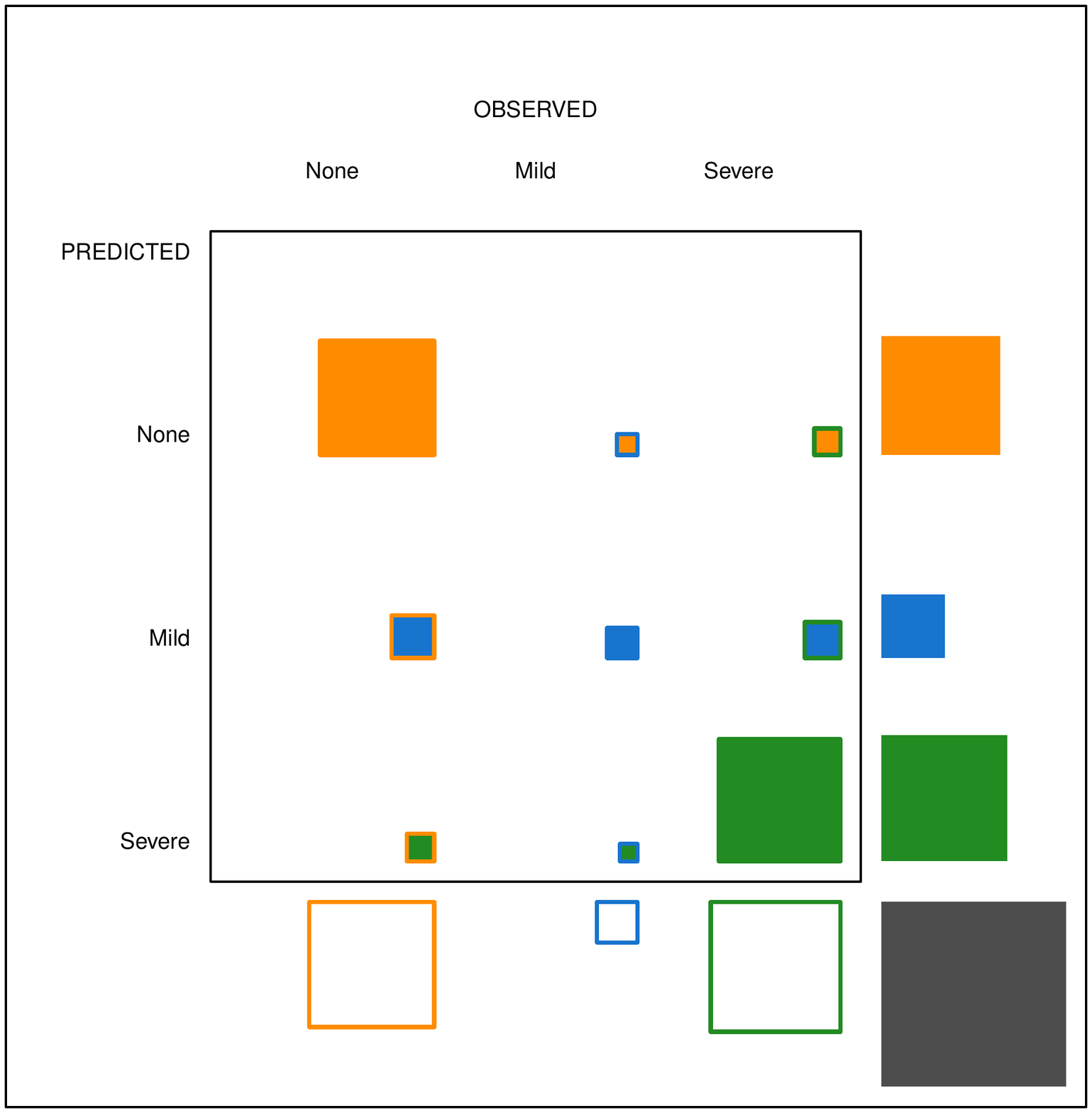}
  \caption{\label{fig:confmat}
          A fluctuation plot to visualize the confusion matrix given in the table \ref{tab:confmat}.}
\end{figure}

\begin{figure}[htb]
\centering
\begin{tabular}{cc}
\includegraphics[width=0.4\linewidth]{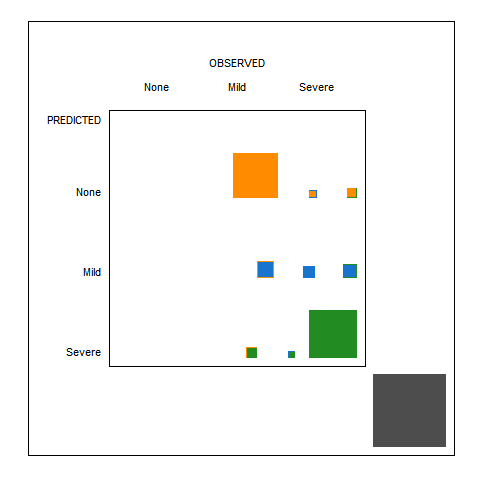}
&\includegraphics[width=0.4\linewidth]{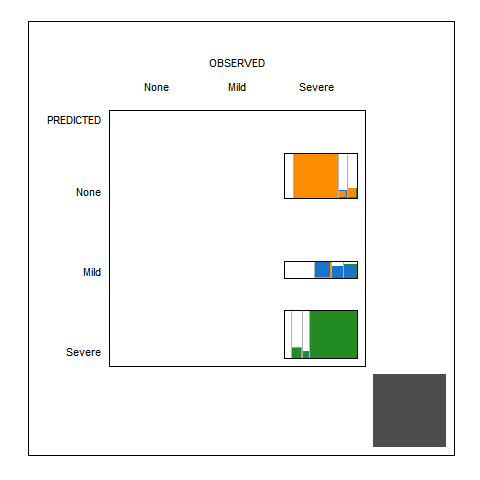}\\
(a) &(b)\\
\includegraphics[width=0.4\linewidth]{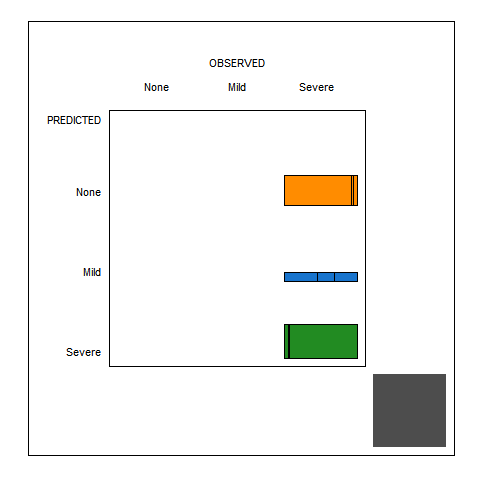}
&\includegraphics[width=0.4\linewidth]{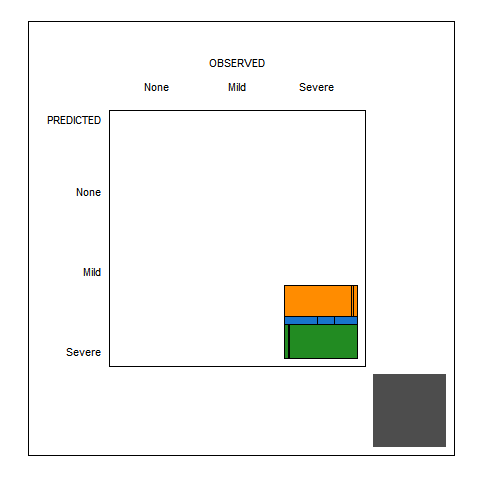}\\
(c) &(d)\\
\includegraphics[width=0.4\linewidth]{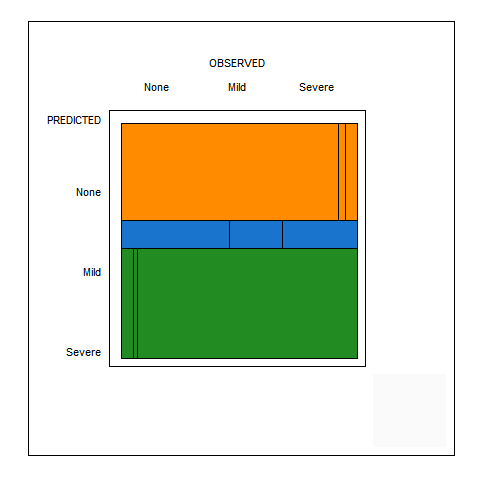}\\
(e)
\end{tabular}
\caption{\label{fig:aquanimmatconf}%
\textit{Aquanim} of the confusion matrix from the fluctuation diagram to the mosaic plot}
\vspace{-.5cm}
\end{figure}

The confusion matrix shown in Table \ref{tab:confmat} displays the number of occurrences of correct and erroneous predictions of the model assessed on an image data set with known ground truth labels provided by human volunteers.

A confusion matrix is a contingency table that  can be read in different ways \cite{WickhamH11}\cite{Friendly_extendingmosaic}.
Each number in a cell counts the data having a specific pair of observed (true) and predicted labels. This number over the total of all numbers in the matrix accounts for the proportion  of such cases that we expect to happen in real conditions when the classifier is used to classify new unseen data. This proportion is an estimator of the joint probability $p(class_o,class_p)$ that $class_o$ is observed and $class_p$ is predicted.  

Classifiers are sensitive to class imbalance. For instance we can read from the table that about $200$ data are from the Mild damage class, while about ten times more are images of no damage and severe damage. If the humanitarian organizations have a specific program to run for each type of damage, it is important that the classifier be reliable for any type. Here we expect it will not be good to detect Mild damages. 

We use a fluctuation diagram to visualize the confusion matrix as in the figure \ref{fig:confmat}. This encodes the numbers as squares whose areas are proportional to the joint probabilities. Squares' fill color codes for the predicted class, and edges' color for the observed class. We also use a fluctuation diagram to display the marginal proportions. The area of the yellow filled square on the top right out of the inner frame map the total amount of images predicted with no damage whatever the ground truth. The area of the gray square at the bottom right account for the total number of images so its area gives the probability unit.

We can decompose the joint probability into the marginal $p(class_p)$  and conditional $p(class_o|class_p)$ probabilities, all related by the Bayes rule: 
$p(class_o,class_p)=p(class_o|class_p)p(class_p)$.
Then we see as studied in depth in \cite{WickhamH11}, that the area of the squares in the fluctuation diagram can be decomposed into the product of a marginal probability and a conditional one. The joint probability can be conditioned  on the predicted classes as above, or on the observed classes:
$p(class_o,class_p)=p(class_p|class_o)p(class_o)$
We can use a mosaic plot to visualize these quantities more efficiently.
The Bayesian rule ensure than the area of corresponding cells in both plots is identical, but the shape of the cells in the mosaic plot are rectangles whose edge length directly encode the marginal and conditional probabilities of interest.

We design an \textit{aquanim} to make clear the relation between the two plots.

\subsection{Animation design}

The animation is displayed in the figure \ref{fig:aquanimmatconf}. We first move all  the squares on the right side of the plot (a) . 
The total area of squares in the first row $Np$ (predicted None) is equal to $p(No,Np)+p(Mo,Np)+p(So,Np)=p(Np)$. Doing the same for each row, we see that summing the area of each square in the right margin equals $p(Np)+p(Mp)+p(Sp)=1$ which is the area and probability of the gray square at the bottom right representing all the cases.

A FAIRE We could use an \textit{aquanim} tip to show that identity equation by transferring the liquids from the 3 marginal squares in the right margin inside the gray square below showing they fill it exactly. The same is true for the ones in the bottom margin.

We use the single-container reshaping block to reshape the 3 squares of the first row into a rectangle with unit length and height $p(Np)$ (b)(c). We do the same for the other two rows. By forcing the height of these rectangles, we force the incompressible liquid say of the left most square to match the $p(Np)$ height and so to spread along this unit length band in proportions $x$ such that:
$p(Np)*x=p(No,Np)$ its original invariant area. By Bayes rule $x=p(No|Np)$ so we get the Bayes decomposition we need for each segment in these bands. 

Each horizontal band is then piled up (d). All three form a square mosaic plot of unit length edges, of same size as  the gray square.
We resize this mosaic plot (e,f) to fit in the full matrix so it is easier to read.

Now two patterns become obvious: the blue class is predicted about ten times less than the other ones. The segments of the blue bar are roughly in proportions $50\%$, $20\%$, and $30\%$, so given the blue Mild class has been predicted, there is only $20\%$ chance that there are actually Mild damages.

\section{Discussion}

We proposed \textit{aquanims} for all types of transitions  described in \cite{HeerR07} except for the \textit{view transformations} like pan or zoom which are global or relative area-preserving respectively, and \textit{substrate transformations} (non-linear scaling) whose generated space distortions make more difficult \textit{aquanims} design. In the latter case, we should define first what area preserving means in these non-euclidean spaces.

We did not explore \textit{aquanims} with treemaps, but our study of mosaic plots involving the need for free space between the tiles showed that designing \textit{aquanims} to animate changes in densely packed treemaps would be challenging. A possibility would be to explode the tiles to get free space between them for reshaping some of them with \textit{aquanims} then collapse them back in final position. It seems possible to design \textit{aquanims} with Voronoi treemaps \cite{Balzer05} as the cells' geometry is less constrained by rigid alignments as in standard treemaps or mosaic plots. In that case however the challenge would be to evoke the hydraulic metaphor  with partially filled Voronoi cells for instance. 

In communicating-containers or communicating-segments blocks (Sections \ref{sec:communicatingcontainerstransfer} and \ref{sec:communicatingsegmentsshift}), the decreasing area of some cells synchronized with the increasing area of other ones so the sum of their areas is invariant, seems to be key. The role of this dynamic feature to support the perception of area-preservation is still to be investigated.
In isolated-container-reshape blocks (Section \ref{sec:isolatedcontainerreshape}), the role of numbers and types of moving edges is also to be investigated.

We also saw that color can be used to code for non area preserving animations in combination with \textit{aquanims} (Section \ref{sec:histoadaptchangingdata}) so color coding compensate for that issue REPRENDRE CETTE PHRASE.

\section{Conclusion}

We presented \textit{aquanims} animations based on a hydraulic metaphor. The first principle of \textit{aquanims} states that counts or proportion data are mapped to liquids volume  instead of containers in area-based  charts. The second principle states that liquids are incompressible supporting the design of \textit{aquanims} as area-preserving animated transitions. We present \textit{aquanims} building-blocks and show how they can be used to design various \textit{aquanims} that avoid occlusions and preserve areas.

How effective are these animations for the user remains to be evaluated. Studying how \textit{aquanims} could be used in treemaps and other non-rectangle area-based  charts is let as a future work.

\subsection{References}
\bibliographystyle{eg-alpha-doi}
\bibliography{egbibsample}

\end{document}